\documentclass[twocolumn,showpacs,preprintnumbers]{revtex4}
\usepackage{amsfonts}
\usepackage{amsmath}
\usepackage{amssymb}
\usepackage{graphicx}
\usepackage{dcolumn}
\usepackage{bm}

\input{tcilatex}

\begin{document}

\preprint{}
\title{Computation of topological charges of optical vortices via
non-degenerate four-wave-mixing}
\author{Wei Jiang}
\email{jwayne@mail.ustc.edu.cn}
\author{Qun-feng Chen}
\author{Yong-sheng Zhang}
\author{G. -C. Guo}
\affiliation{Key Laboratory of Quantum Information, University of Science and Technology
of China, Hefei, 230026,P. R. China}

\begin{abstract}
In this paper, we report an experiment, which demonstrates computation of
topological charges of two optical vortices via non-degenerate
four-wave-mixing process. We show that the output signal beam carries
orbital angular momentum which equals to the subtraction of the orbital
angular momenta of the probe light and the backward pump light. The $^{85}Rb$
atoms are used as the nonlinear medium, which transfer the orbital angular
momenta of lights.
\end{abstract}

\maketitle

Although polarization of light, which depends on spin angular momentum of
photon, is familiar to people, its closely related counterpart, orbital
angular momentum (OAM) of light, was quite unfamiliar to the researchers
until very recently\cite{Allen}. The most common light beam that carries OAM
is the so called Laguerre-Gaussian (LG) mode beam\cite{Kogelnik}. Optical
beams with OAMs include screw topological wave front dislocation or
vortices. A spiral phase ramps around a singularity where the phase of the
wave is undefined. The order of the singularity multiplied by its sign is
referred as the topological charge of the dislocation. This kind of light
field is also called optical vortex (OV) due to its helicoidal wavefront\cite%
{Cullet}.

There is a great deal of interest in the property of light with OAM and how
it interacts with materials. Transfer of OAM from light to macroscopic
particles has been reported\cite{He, He1, Simpson}. Using an light beam with
OV to excite vortex state in a Bose-Einstein condensation (BEC) was proposed%
\cite{Marzlin, Anglin}. Experiments of transfer of OAM between light and
atoms were reported\cite{Tabosa, Tabosa1, Tabosa2, Kozuma}. Another
interesting direction is studing the change of OAM of light during a
nonlinear optical process. Conversion of topological charge of optical
vortices in a Second-Harmonic Generation (SHG) has been reported\cite%
{Bassistiy, Padgett, Allen1, Padgett1}. Studies of OAM of light in
parametric down-conversion processes have also been explored\cite{Zeilinger,
Arnaut, Arlt}.

In this paper we report an experiment, which demonstrates computation of
topological charges of two optical vortices via non-degenerate
four-wave-mixing. $^{85}Rb$ atoms are used to mediate the interaction among
lights with optical vortices. Fig. 1 shows the experimental setup. The 3cm
vapor cell contains isotopically pure $^{85}Rb$. The temperature of the cell
is kept around 60$^{\circ }$C, corresponding to a atomic density about 3.5$%
\times 10^{11}$cm$^{-3}$. A three layer magnetic shield is used to screen
out the magnetic field of earth. The remaining magnetic field inside the
shield is less than 1mGs. Three external cavity diode lasers (ECDL) are used
in our experiment. Two ECDLs working at 795nm are used as the forward pump
field and the probe field. A 780nm ECDL serves as the backward pump field.
The frequency of the forward pump field is tuned to be resonant with $%
\left\vert 5S_{1/2},F=2\right\rangle \rightarrow \left\vert
5P_{1/2},F=2\right\rangle $ transition. The frequency of the backward pump
field is tuned to be resonant with $\left\vert 5S_{1/2},F=3\right\rangle
\rightarrow \left\vert 5P_{3/2},F=2\right\rangle $ transition. And the
frequency of the probe field is tuned to be resonant with $\left\vert
5S_{1/2},F=3\right\rangle \rightarrow \left\vert 5P_{1/2},F=2\right\rangle $
transition. The forward pump field is horizontally polarized. The backward
pump field and the probe field are vertically polarized. The backward pump
field and the probe field are focused inside the cell. The diameters of them
are about 0.5mm. The forward pump beam is collimated and has a diameter
about 3mm. The forward pump field and the backward pump field are made to be
counter-propagating. There is a small angle (about 10mrad) between the probe
field and the two pump fields. The three beams are made to be coincided
inside the vapor cell. This kind of configuration forms a closed
non-degenerate four-wave-mixing (FWM) process, which produces a 780nm
signal. And the configuration of the polarizations ensures a maximum output
signal with horizontal polarization. The output FWM signal is picked up by a
PBS and then sent into a Mach-Zender (M-Z) interferometer. This
interferometer serves as an analyzer of the OAM of the output light field.
We will explain it in detail later. The output of the M-Z interferometer is
monitored by a CCD camera.

We use two computer generated holograms to make the probe field and the
backward field to be LG mode beams, which carry OAMs (or topological charges)%
\cite{Mask}. These holograms have fork-like patterns in their center (see
Fig. 1). When the hologram is illuminated by a normal Gaussian laser beam,
the first-order diffracted beams will carry OAMs of $+\hbar $ or $-\hbar $
depending on the sign of the diffraction order. The field amplitude of a LG
mode laser is given by\cite{Allen},

\begin{equation}
E_{lp}^{LG}=E_{0}(\frac{\sqrt{2}r}{w})^{l}e^{-il\phi }e^{-\frac{r^{2}}{w^{2}}%
}L_{p}^{l}(\frac{2r^{2}}{w}),  \label{lg}
\end{equation}%
where, $E_{0}$ is the amplitude. $w$ is the half-beam width. $L_{p}^{l}$ is
the Laguerre polynomial. $r$ and $\phi $ are, respectively, the radial and
angular coordinates in cylindrical polar coordinate system with its $z$ axis
being along the beam propagation direction. In our experiment all three
light fields are output from single mode fibers and are in TEM$_{00}$ modes.
We put two holograms after the output couplers of the backward pump beam and
the probe beam respectively. The first-order diffraction is used. We use two
translation stages to control the transverse position of the holograms. If
the fork-shape structure is illuminated, the diffracted beam is a $%
LG_{0}^{1} $ beam, which carries $\pm \hbar $ of OAM. The sign of the OAM
depends on the order of the diffraction ($+1$ or $-1$). If we move the
fork-shape structure outside the beam, then the diffracted beam is an
ordinary Gaussian beam, which carries no OAM. The diffraction efficiencies
of the two holograms are about 10\%, which is quite low. The power of
forward pump field, backward pump field and probe field are all about 700$%
\mu $W.

Before presenting the experimental results, let's first review some basic
properties of the FWM process. This process utilizes the third-order
nonlinearity $\chi ^{(3)}$ of the nonlinear medium. The output FWM signal is
related with the input fields with relation\cite{Boyd},%
\begin{equation}
E_{S}=\chi ^{(3)}E_{F}E_{B}E_{P}^{\ast },  \label{fwm}
\end{equation}%
where $E_{S}$, $E_{F}$, $E_{B}$ and $E_{P}$ are electrical amplitudes of the
signal field, the forward pump field, the backward pump field and the probe
field respectively. Here we assume scalar fields for simplicity.

The energy conservation law and the momentum conservation law must be
obeyed. These conditions suggest that $\omega _{F}+\omega _{B}-\omega
_{P}=\omega _{s}$ and the phase-matching condition $\vec{k}_{F}+\vec{k}_{B}-%
\vec{k}_{P}=\vec{k}_{S}$. In the traditional FWM experiment all beams carry
no OAM. So the conservation of OAM is automatically fulfilled. But in our
experiment, since we introduce OAMs into the backward pump field and the
probe field, the OAM conservation law must be considered too. To write it
explicitly,%
\begin{equation}
l_{F}+l_{B}-l_{P}=l_{S}.
\end{equation}

First we only make the backward pump field to carry OAM (i.e. $l_{B}=1$).
Since $l_{F}=l_{P}=0\,$, the output FWM signal should also carries $+\hbar $
of OAM (i.e. $l_{S}=1$). To analyze the OAM of the output beam, A M-Z
interferometer is used. This M-Z interferometer is different from the
ordinary one. The beam is reflected odd times in one arm and is reflected
even times in the other. The reason we use this kind of interferometer is
that reflection can flip the sign of the LG beam\cite{reflect}. A LG beam
which carries $+l\hbar $ of OAM is changed to a LG beam which carries $%
-l\hbar $ of OAM after reflection. So the beams from the two arms of the M-Z
interferometer have OAMs with opposite signs. When they interfere, they will
produce an interference pattern that can be used to reveal the OAM carried
by the input field. Fig. 2 (a) and (b) are the images recorded by CCD
camera. (a) is the recorded signal when one arm of the M-Z interferometer is
blocked. The field distribution is a doughnut-shape spot. Fig. 2(b) is
recorded without the block. This time the signal fields from the two arms
form an two-part spiral interference pattern. From these two images we know
that the signal beam carries OAM of $+\hbar $. One can easily reproduce this
pattern using Eq.(\ref{lg}). The field at the CCD camera is,%
\begin{equation}
E(r,\phi )\varpropto E_{0}(\frac{\sqrt{2}r}{w})e^{-\frac{r^{2}}{w^{2}}%
}L_{0}^{1}(\frac{2r^{2}}{w})(e^{-i\phi }+e^{i\phi }e^{i\varphi })e^{i\eta r},
\label{inter}
\end{equation}%
where $\eta $ is a parameter which represents the difference of the
divergences of the two beams in the two arms of the M-Z interferometer. $%
\varphi $ is the phase factor caused by the different optical lengths. This
term causes the interference pattern to rotate. The two-part structure is
caused by term $(e^{-i\phi }+e^{i\phi }e^{i\varphi })$. The spiral shape is
caused by the last term in Eq.(\ref{inter}). In the experiment we make the
divergences of the beams in the two arms of the M-Z interferometer different
deliberately. The consequence is the spiral shape interference pattern,
which can be recognized more easily. Fig. 2(c) and (d) is the result of
theoretical calculation using Eq.(\ref{inter}). We can see that they are in
good agreement with the experimental results.

After this first step we carry out an experiment, in which two beams with
topological charge are used. The topological charge of the backward pump
beam is $+1$($l_{B}=+1$). The probe beam has the same topological charge as
the backward pump beam. But it is reflected by a mirror before entering the
vapor cell, so as mentioned above the reflection flips the sign of the
topological charge. Therefore it carries $-\hbar $ of OAM ($l_{P}=-1$). From
the angular momentum conservation law we know $%
l_{S}=l_{F}+l_{B}-l_{P}=0+1-(-1)=2$. We also analyze the output FWM signal
with the M-Z interferometer. The results are shown in Fig. 3. Fig.3 (a) is
the recorded signal when one arm of the M-Z interferometer is blocked. The
field distribution is also a doughnut-shape spot. One thing worth noting is
that the internal circle is actually not a circle but an irregular shape.
This phenomena is caused by the geometry of the experimental setup. Because
in our setup the probe beam and the two pump beams are not collinear
exactly. There is a small angle (10mrad) between them. So they only coincide
partly inside the cell. Consequently the output signal is not a perfect
doughnut shape. Fig. 2(b) is recorded without the block. The interference
pattern is a four-part structure, which has a windmill shape. These two
images manifest the OAM of the output beam is $2\hbar $. In this sense this
experiment can be viewed as an computation of the topological charges of
backward pump field and the probe field. In our case, it is a subtraction.

This experiment demonstrates that OAM is an intrinsic property of light,
just as polarization. In a nonlinear process which involves lights with
non-zero OAM, the OAM conservation law must be obeyed also.

One can easily extend this setup to compute three topological charges by
introducing OAM in the forward pump field. In principle, N-1 topological
charges can be mixed using a N-wave-mixing process. However higher-order
nonlinear process has a much smaller cross section. And cascading of many
FWM processes seems to be a more feasible way to do computation of OAMs of
lights. Compared with other experimental schemes, such as SHG\cite%
{Bassistiy, Padgett, Allen1, Padgett1}, the FWM configuration provides a
larger flexibility.

In conclusion, we have reported an experiment which realized an computation
of topological charges of two optical vortices. This experiment utilize the
OAM conservation nature of the FWM process. The $^{85}Rb$ atoms are used as
the nonlinear medium, which transports the OAMs of lights. We show that OAM
can be transferred from one beam to another. Then we demonstrate the
subtraction of OAMs of two light beams. This experiment may find
applications in optical computing and processing.

\begin{acknowledgments}
We thank X. -F. Ren for providing the holograms and some helpful
discussions. This work was funded by National Fundamental Research Program
(2001CB309300), National Natural Science Foundation of China (Grant No.
60121503, 10304017), the Innovation funds from Chinese Academy of Sciences,
and Program for New Century Excellent Talents in University.
\end{acknowledgments}

\bigskip

\textbf{Figure Captions}

Fig. 1 (a) Experimental setup. P: Polarizer; BS: Beam splitter; PBS:
Polarized beam splitter. The 3cm vapor cell contains isotopically pure $%
^{85}Rb$. A three layer magnetic shield is used to screen out the magnetic
field of earth. Three external cavity diode lasers (ECDL) are used in our
experiment. The backward pump field and the probe filed are focused inside
the cell. The diameters of them are about 0.5mm. The forward pump beam is
collimated and has a diameter about 3mm. The forward pump field and backward
filed are made to be counter-propagating. There is a small angle (about
10mrad) between the probe field and the two pump fields. The three of them
are made to be coincided inside the vapor cell. The output FWM signal is
picked up by the PBS and then sent to a Mach-Zender (M-Z) interferometer.
One output of the M-Z interferometer is directed to a CCD camera. This
interferometer serves as an analyzer of the OAM of the incident beam. (b)
Energy diagram of $^{85}Rb$. The frequency of the forward pump field is
tuned to be resonant with $\left\vert 5S_{1/2},F=2\right\rangle \rightarrow
\left\vert 5P_{1/2},F=2\right\rangle $ transition. The frequency of the
backward pump field is tuned to be resonant with $\left\vert
5S_{1/2},F=3\right\rangle \rightarrow \left\vert 5P_{3/2},F=2\right\rangle $
transition. And the frequency of the probe field is tuned to be resonant
with $\left\vert 5S_{1/2},F=3\right\rangle \rightarrow \left\vert
5P_{1/2},F=2\right\rangle $ transition. The forward pump field is
horizontally polarized. The backward pump field and the probe field are
vertically polarized.

Fig. 2 (a) Recorded signal when one arm of the M-Z interferometer is
blocked. The field distribution is a doughnut-shape spot. (b) Recorded
signal without the block. This time the signal fields from the two arms
forms an two-part spiral interference pattern. (c) and (d) is the result of
theoretical calculation using Eq.(\ref{inter}).

Fig. 3 The same as Fig. 2, except that two input beams (the backward pump
beam and the probe beam) contain non-zero OAM. The OAM of the output signal
equals to the subtraction of the OAM of the backward pump field and the OAM
of the probe field.

\end{document}